\documentclass{ifacconf}

\usepackage{graphicx} 
\usepackage[cmex10]{amsmath} 
\usepackage{amssymb}  
\usepackage{natbib}        
\usepackage{microtype,enumerate,url}
\usepackage{pgfplots}
\usepackage[bb=boondox,bbscaled=.95,cal=boondoxo]{mathalfa}
\usepackage{tikz} 
\usetikzlibrary{patterns,calc,decorations.pathmorphing,decorations.markings} 
\renewcommand{\theenumi}{(\roman{enumi})}

\newtheorem{theorem}{Theorem}

\newtheorem{example}{Example}

\newtheorem{definition}{Definition}

\newcommand{\beq}{\begin{equation}}
\newcommand{\eeq}{\end{equation}}
\newcommand {\bseq}{\begin{subequations}}
\newcommand {\eseq}{\end{subequations}}
\newcommand {\bma}{\left[}
\newcommand {\ema}{\right]}

\newcommand {\R}{\mathbb{R}} 	
\newcommand {\Rplus}{\mathbb{R}_{+}} 	
\newcommand {\Co}{\mathbb{C}} 	
\newcommand {\Coplus}{\Co_{+}} 	
\renewcommand{\Re}{\mathbf{Re}} 
\newcommand{\transpose}{\mathsf{T}} 
\newcommand{\norm}[1]{\left\lVert#1\right\rVert}

\newcommand{\overbar}[1]{\mkern 1.5mu\overline{\mkern-1.5mu#1\mkern-1.5mu}\mkern 1.5mu}

\newcommand{\TAC}{\textit{{IEEE} Trans. Autom.
  Control}}\newcommand{\SCL}{\textit{Syst. Control
  Lett.}}  

\begin{document}

\begin{frontmatter}

\title{Dominance margins for feedback systems\thanksref{footnoteinfo}} 

\thanks[footnoteinfo]{\noindent The research leading to these results has received funding from the European Research Council under the Advanced ERC Grant Agreement Switchlet n. 670645.}

\author{Alberto Padoan, Fulvio Forni, Rodolphe Sepulchre} 

\address{Department of Engineering, University of Cambridge,\\ 
							Cambridge, CB2 1PZ,  UK \\ 
							(Email: \{ {\tt \small a.padoan |  f.forni | r.sepulchre \} @eng.cam.ac.uk} ) }

\begin{abstract}  
\noindent 
The paper introduces notions of robustness margins geared towards the analysis and design of systems that switch and oscillate. While such phenomena are ubiquitous in nature and in engineering, a theory of robustness for behaviors away from equilibria is lacking. The proposed framework addresses this need in the framework of $p$-dominance theory, which aims at  generalizing stability theory for the analysis of systems with low-dimensional attractors. Dominance margins are introduced as natural generalisations of stability margins in the context of $p$-dominance analysis. In analogy with stability margins, dominance margins are shown to admit simple interpretations in terms of familiar frequency domain tools and to provide quantitative measures of robustness for multistable and oscillatory behaviors in Lure systems. The theory is illustrated by means of an elementary mechanical example. 
\end{abstract}


\end{frontmatter}

\section{Introduction}

Robustness is a classical concept in engineering. In broad terms, a behavior  is robust if it persists under the effect of exogenous perturbations or parametric uncertainty.  One of the key questions of control theory is to understand when the \textit{stability} of a system is robust to model uncertainty~\citep{bode1945network,francis1987course,mcfarlane1990robust,doyle1992feedback,green1995linear,zhou1996robust,vinnicombe2001uncertainty,astrom2008feedback}. For linear time-invariant systems, the classical notions of \textit{gain margin} and \textit{phase margin} provide quantitative measures of robustness that remain the basis of  control engineering practice to date~\citep{astrom2008feedback}. These notions possess an intuitive interpretation in terms of Nyquist and Bode diagrams. They can be extended to nonlinear systems using the notion of \textit{disk margin}~\citep{sepulchre1997constructive} and they are at the root of modern robust control theory.

The paper introduces notions of robustness margins geared towards the analysis and design of systems with multistable and oscillatory behaviors, \textit{i.e.}  systems with  multiple stable equilibria and simple attractors. These behaviors are of great importance in electronics~\citep{chua1987linear} and mechanics~\citep{guckenheimer1991nonlinear} and are also believed to lie at the heart of key biological questions, including the generation of spikes in the brain~\citep{izhikevich2007dynamical} and the design of synthetic genetic circuits~\citep{del2018future}. This goal is pursued by adopting a \textit{differential} viewpoint, \textit{i.e.} using the linearization of a system along arbitrary solutions to infer global properties of the system, and shifting the focus from stability analysis to \textit{$p$-dominance} analysis~\citep{forni2018differential,felix2018analysis}, \textit{i.e.} convergence to $p$-dimensional attractors rather than to an equilibrium point.  
  

Dominance margins are introduced as natural generalizations of stability margins in the context of $p$-dominance analysis.  In analogy with stability margins, these notions are shown to admit simple interpretations in terms of familiar frequency domain tools and to provide practical measures of robustness for multistable and oscillatory behaviors  in Lure systems, \textit{i.e.} systems that can be described as the feedback interconnection of a linear, time-invariant system and a locally Lipschitz function.  The significance of these notions for analysis and design is demonstrated studying the robustness of multistability and oscillations in a mechanical system using a saturated feedback law.

The remainder of the paper is organized as follows. 
Section~\ref{sec:example} describes the motivating mechanical example. Section~\ref{sec:preliminaries} recalls some preliminary results on $p$-dominance theory from~\citep{forni2018differential} and~\citep{felix2018analysis}. 
Section~\ref{sec:main-results} contains the main results of the paper,  where notions of gain margin, phase margin, and disk margin are defined for $p$-dominant systems.  Section~\ref {sec:discussion} provides an outlook to future research directions.
Section~\ref{sec:conclusion}  summarises the main results of the paper.

\textbf{Notation}
$\R$, $\R^n$ and $\R^{p \times m}$  denote the set of real numbers, the set of $n$-dimensional real vectors and the set of $p \times m$-dimensional real matrices, respectively. 
$\Rplus$ and $\Coplus$ denote the set of non-negative real numbers and the set of 
complex numbers with non-negative real part. $j$ denotes the imaginary unit and $j\R$ denotes the set of complex numbers with zero real part.
$I$ denotes the identity matrix.
$M^{\transpose}$ denotes the transpose of the matrix ${M\in\R^{p\times m}}$.
$D(K_1, K_2)$  denotes the closed disk in the complex plane whose diameter connects the points $-1/K_1$ and $-1/K_2$ for ${K_1 \in\R}$ and ${K_2 \in\R}$, with ${K_1 < K_2}$ and ${K_1 K_2 > 0}$; by a convenient abuse of notation, if ${K_1 K_2 < 0}$ then $D(K_1, K_2)$  denotes the complement of the closed disk with its center on the real axis and its boundary intersecting the real axis at the points $-1/K_1$ and $-1/K_2$; if ${K_1 = 0}$ or ${K_2 = 0}$  then $D(K_1, K_2)$ denotes the open half-plane to the left or to the right of the lines defined by ${\Re(s) = - 1/K_2}$ or ${\Re(s) = - 1/K_1}$, respectively. In all these cases, $D(K_1, K_2)$ is referred to as a disk. 

\section{A motivating example} \label{sec:example}

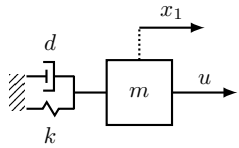
\begin{figure}[h]
\centering

 \tikzstyle{spring}=[thick,decorate,decoration={zigzag,pre length=0.2cm,post
 length=0.2cm,segment length=6}]

 \tikzstyle{damper}=[thick,decoration={markings,  
   mark connection node=dmp,
   mark=at position 0.5 with 
   {
     \node (dmp) [thick,inner sep=0pt,transform shape,rotate=-90,minimum
 width=10pt,minimum height=2pt,draw=none] {};
     \draw [thick] ($(dmp.north east)+(1.25pt,0)$) -- (dmp.south east) -- (dmp.south
 west) -- ($(dmp.north west)+(1.25pt,0)$);
     \draw [thick] ($(dmp.north)+(0,-2.5pt)$) -- ($(dmp.north)+(0,2.5pt)$);
   }
 }, decorate]

 \tikzstyle{GM}=[fill,pattern=north east lines,draw=none,minimum width=0.55cm,minimum height=0.2cm]
 
 \begin{tikzpicture}[color=black, scale=.86, every node/.style={transform shape}]

 \node[draw,outer sep=0pt,thick][minimum width=1cm, minimum height=1cm] (M1) at (5.25,-1.3) {$m$};

\node (v1) at ($(M1)-(1,0)$){}; 
\node (GM1) [GM,anchor=north,rotate=90] at ($(v1) - (1,0)$) {};
\draw[thick]($(M1.west)$) -- ($(v1)+(0,0)$) {};
\draw[thick]($(v1)-(0,-0.26)$) -- ($(v1)-(0,0.26)$) {}; 
\draw[spring] ($(v1)-(0,0.25)$) --  ($(GM1)-(-.15,0.25)$)  node [midway, below=.15cm] {$k $};
\draw[damper] ($(v1)-(0,-0.25)$) --  ($(GM1)-(-.15,-0.25)$)  node [midway, above=.275cm] {$d$}; 

\draw[thick, densely dotted] ($(M1.north)$) -- ($(M1.north) + (0,.5)$);
\draw[thick, -latex] ($(M1.north) + (0,0.5)$) -- ($(M1.north) + (1,0.50)$) node [midway, above] {$x_1$};

 \draw[thick, latex-] ($(M1.east) + (1,0)$) -- ($(M1.east)$) node [midway, above] {$u$};

\end{tikzpicture}
\centering
\caption{{The controlled mass-spring-damper system~\eqref{eq:system-MSD}.}}
\label{fig:case-study}
\end{figure}

Consider a linear, time-invariant, mass-spring-damper system   described by the equations 
\beq \label{eq:system-MSD}
\dot{x}_1 = x_2,  \quad
m\dot{x}_2 = - k x_1-dx_2+u , \quad
y = x_1,
\eeq
in which ${x_1(t) \in \R}$ is the position of the point mass, ${x_2(t) \in \R}$ is the velocity of the point mass,  ${u(t)\in \R}$ is the exogenous (force) input, ${y(t) \in \R}$ is the measured (position) output,  ${m \in \Rplus}$ is the mass of the point mass,  ${k \in \Rplus}$ is the spring constant,  and ${d \in \Rplus}$ is the damping coefficient, respectively.  
Fig.~\ref{fig:case-study} provides a diagrammatic representation of system~\eqref{eq:system-MSD}.
The parameters are selected as  ${m = 1 \, \mathrm{kg}}$, ${k = 1 \,\mathrm{N} \cdot \mathrm{m}^{-1}}$, ${d = 5 \, \mathrm{kg}\cdot\mathrm{s}^{-1}}$.

\begin{figure}[h!]
\centering
\includegraphics[trim={0cm 6.7cm 0cm 0.5cm}, 
								clip ,
								width=\columnwidth]{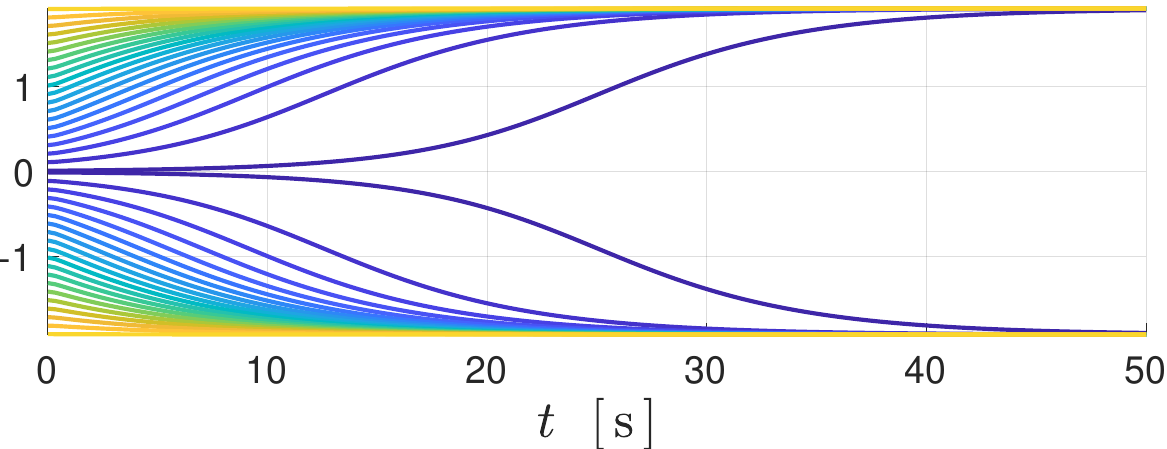}
\caption{Time history of the output of system~\eqref{eq:system-MSD} for different initial conditions with input~\eqref{eq:controller-P}, with   ${k_P = -2}$.}
\label{fig:msd_time_bistability}
\end{figure}

Consider the question of designing a controller that renders the system~\eqref{eq:system-MSD}
bistable via feedback. A simple solution  is provided by  the nonlinear {\it proportional} feedback controller
\beq \label{eq:controller-P}
u = - k_P \tanh(y), \quad  k_P \in \R.
\eeq
For \textit{negative} values of $k_P$ the feedback law~\eqref{eq:controller-P} describes a 
\textit{positive} feedback loop. For sufficiently large values of ${k_P<0}$, the stable equilibrium at the origin of the closed-loop system undergoes a supercritical  pitchfork bifurcation and a (globally) bistable behavior emerges, as illustrated in Fig.~\ref{fig:msd_time_bistability}. 

\begin{figure}[h!]
\centering
\includegraphics[trim={0cm 6.7cm 0cm 0.5cm}, 
								clip ,
								width=\columnwidth]{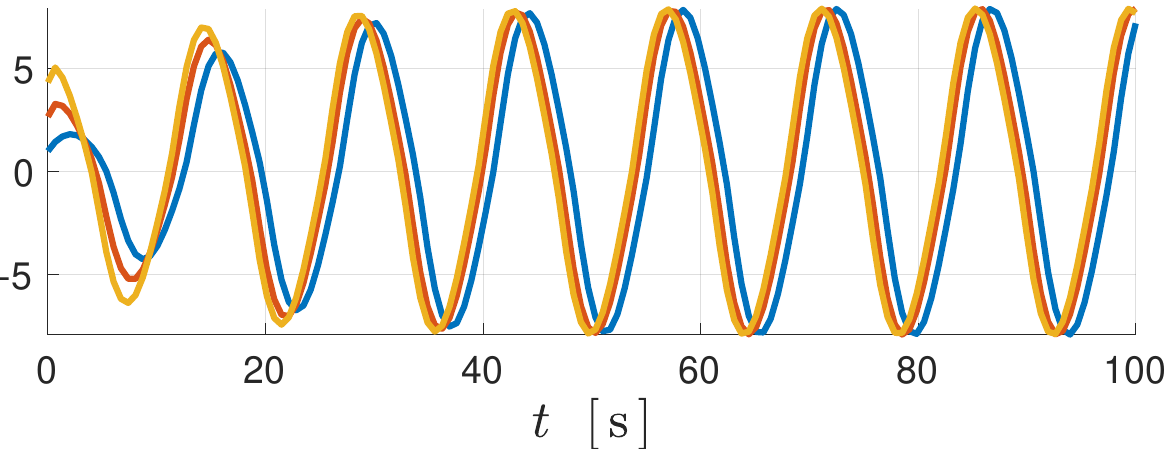}
\caption{Time history of the output of system~\eqref{eq:system-MSD} for different initial conditions with input~\eqref{eq:controller-PI}, with   ${k_I = -5}$.}
\label{fig:msd2_time_oscillations}
\end{figure}

Similarly, consider the question of design a controller that renders the system~\eqref{eq:system-MSD}
oscillatory via feedback. A simple solution is the nonlinear {\it integral} feedback controller
\beq \label{eq:controller-PI}
u = - k_I \tanh\left(\int y \right), \quad k_I \in \R .
\eeq
For sufficiently small values of ${k_I<0}$, the slow adaptation loop provides a feedback mechanism for a global oscillation~\citep{sepulchre2005feedback}, as illustrated in Fig.~\ref{fig:msd2_time_oscillations}.

We consider these  design problems  for simplicity. The combination of~\eqref{eq:system-MSD},~\eqref{eq:controller-P} and~\eqref{eq:controller-PI}, however, can be regarded as the source of bistability  and oscillations  in  other classical examples. For example,  the Duffing oscillator and the pendulum~\citep[p.82 and p.56]{guckenheimer1991nonlinear} can be are obtained by substituting the nonlinearity in~\eqref{eq:controller-P} with a cubic nonlinearity and with a sinusoidal nonlinearity defined on the unit circle, respectively.

The purpose of this example is to study the  \textit{robustness} of the behaviors  induced by the controllers~\eqref{eq:controller-P} and~\eqref{eq:controller-PI}. For example, what is the maximal admissible variation in ${k_P}$ and ${k_I}$ which does not affect the  behavior  of the closed-loop system?  Do these controllers   work in the presence of unmodelled dynamics?  

The classical answer to such questions relies on bifurcation theory and is local  in nature. The present paper describes a framework which answers these questions \textit{globally} by combining bifurcation theory with  familiar frequency domain tools. To this end, notions of dominance margins are introduced exploiting existing results from $p$-dominance theory~\citep{forni2018differential,felix2018analysis}, which we recall in the next section for completeness.

\section{Preliminaries} \label{sec:preliminaries}

Consider a continuous-time, nonlinear, time-invariant system described by the equation
\beq \label{eq:system-nonlinear}
\quad \dot{x} = f(x),
\eeq
in which ${x(t)\in\R^n}$ and ${f:\R^{n} \to \R^{n}}$ is a continuously differentiable vector field. The \textit{prolonged system} associated with system~\eqref{eq:system-nonlinear} is defined as
\beq \label{eq:system-nonlinear-prolonged}
\quad \dot{x} = f(x), \quad \delta\dot{x} = \partial f(x) \delta x,
\eeq
with ${x(t)\in\R^n}$, ${\delta x(t)\in \R^n}$ (identified with the tangent space of $\R^n$),
and ${\partial f}$ is the Jacobian of the vector field $f$. 

\begin{definition} \label{def:dominant-linear}
The system~\eqref{eq:system-nonlinear} is $p$-dominant with rate ${\lambda \in \Rplus}$ if there exist ${\varepsilon \in \Rplus}$ and a symmetric matrix ${P\in\R^{n\times n}}$, with inertia\footnote{\!\!
The inertia of the matrix ${A \in \R^{n \times n}}$ is defined as $(\pi,\nu,\delta)$, where $\pi$ is the number of eigenvalues of $A$ in the open right half-plane, $\nu$ is the number of eigenvalues of $A$ in the open left half-plane, and $\delta$ is the number of eigenvalues of $A$ on the imaginary axis, respectively.  
} $(n-p, p, 0)$, such that the prolonged system~\eqref{eq:system-nonlinear-prolonged} satisfies
\beq 
\bma
\begin{array}{c}
\delta \dot{x} \\
\delta x
\end{array}
\ema^{\transpose}
\bma
\begin{array}{cc}
0 & P \\
P &  2\lambda P + \varepsilon I
\end{array}
\ema
\bma
\begin{array}{c}
\delta \dot{x} \\
\delta x
\end{array}
\ema \le 0 
\eeq 
for every $(x,\delta x) \in \R^n \times \R^n$. The property is strict if $\varepsilon >0$.
\end{definition}

The behavior of a $p$-dominant system is characterized by $p$ dominant modes and $n-p$ transient modes. As a result,  the asymptotic behavior is $p$-dimensional, which is particularly significant for small values of $p$ as shown by the following theorem.

\begin{theorem}  \label{thm:asymptotic}
Assume system~\eqref{eq:system-nonlinear} is strictly $p$-dominant with rate ${\lambda \in \Rplus}$. Then every bounded solution of~\eqref{eq:system-nonlinear} converges asymptotically~to 
\begin{itemize}
\item the unique equilibrium  point if ${p=0}$,
\item a (possibly non-unique) equilibrium  point if ${p=1}$,
\item a simple attractor if ${p=2}$, \textit{i.e.} an equilibrium  point, a set of equilibrium  points and their connected arcs or a limit cycle.
\end{itemize}
\end{theorem}

The property of $p$-dominance can be also characterized in the frequency domain~\citep{felix2018analysis}. Consider a continuous-time, single-input, single-output, linear, time-invariant system described by the equations
\beq \label{eq:system-linear-SISO}
\dot{x} = Ax +Bu, \quad y = Cx + Du, 
\eeq
in which ${x(t)\in\R^n}$, ${u(t)\in\R}$, ${y(t)\in\R}$, and ${A\in\R^{n \times n}}$, ${B\in\R^{n \times 1}}$, ${C\in\R^{1\times n}}$ and ${D\in\R}$ constant matrices, with transfer function ${W(s) = C(sI-A)^{-1}B +D}$. For ${\lambda \in \Rplus}$, the corresponding \textit{$\lambda$-shifted transfer function} is defined as
${W_\lambda (s) = W (s-\lambda)}$.

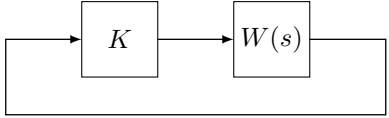
\begin{figure}[h!]
\centering 
\begin{tikzpicture}[scale=1, every node/.style={transform shape}]
\draw  (10,8.5) rectangle (11,7.5);
\node at (10.5,8) {$W(s)$};
\draw  (8,8.5) rectangle (9,7.5);
\node at (8.5,8) {$K$};
\draw[-latex, line width = .5 pt] (11,8) -- (12,8) -- (12,7) --  (7,7) --  (7,8)  -- (8,8);
\draw [-latex,line width = .5 pt](9,8) -- (10,8);
\end{tikzpicture} 
\centering
\caption{Feedback control system.}
\label{fig:robust-1}%
\end{figure}%

\begin{theorem}[Nyquist criterion for $p$-dominance]  \label{thm:nyquist}
Let $W$ \linebreak  be the transfer function of a strictly  $p_1$-dominant with rate ${\lambda \in \Rplus}$ and let ${K\in\R}$, with ${K\not = 0}$. Then the closed-loop system (Fig.~\ref{fig:robust-1})  is strictly $p_2$-dominant with rate $\lambda$ if and only if the Nyquist diagram of $ W_{\lambda}$ encircles  ${(p_2-p_1)}$ times the point $-1/K$  in the clockwise direction.
\end{theorem}

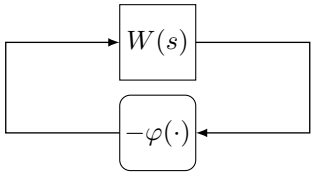
\begin{figure}[h!]
\centering 
\tikzstyle{sum} = [draw, circle, node distance=1cm]
\usetikzlibrary{arrows}
\begin{tikzpicture}
\draw  (-1,3.5) rectangle (0,2.5);
\node at (-0.5,3) {$W(s)$};
\draw[rounded corners]  (-1,2.3) rectangle (0,1.3);
\node at (-0.5,1.8) {$-\varphi(\cdot)$};
\draw[-latex, line width = .5 pt] (0,3) -- (1.5,3) -- (1.5,1.8) -- (0,1.8);
\draw[-latex, line width = .5 pt] (-1,1.8) -- (-2.5,1.8) -- (-2.5,3) --(-1,3) ;
\end{tikzpicture}
\centering
\caption{The Lure system~\eqref{eq:system-Lure}.}
\label{fig:lure}
\end{figure}

A graphical criterion for $p$-dominance can be also established for \textit{Lure systems}, which
can  be  modelled as the negative feedback interconnection of system~\eqref{eq:system-linear-SISO} and a continuously differentiable function\footnote{Local Lipschitz continuity would be sufficient -- continuous differentiability is assumed to streamline the exposition.} ${\varphi:\R \to \R}$ and can be described by the equations 
\beq \label{eq:system-Lure}
\dot{x} = Ax +Bu, \quad y = Cx + Du,  \quad u =  -   \varphi(y).
\eeq
Fig.~\ref{fig:lure} provides a diagrammatic representation of~\eqref{eq:system-Lure}.

The function $\varphi$  is assumed to satisfy the \textit{differential sector condition} ${\partial \varphi \in  [K_1, K_2]}$, defined as
\beq \label{eq:sector-condition} 
(\partial \varphi (y) \delta y - K_1 \delta y) (\partial \varphi (y) \delta y - K_2 \delta y) \leq 0, \  \forall \, y \in \R,
\eeq
in which ${K_1 \in\R}$ and ${K_2 \in\R}$, with ${K_1 < K_2}$. The following statement formalises a circle criterion for $p$-dominance.

\begin{theorem}[Circle criterion for $p$-dominance] \label{thm:circle_criterion}
Consider \linebreak system~\eqref{eq:system-Lure} and let $\lambda \in \Rplus$. Assume 
\begin{itemize}
\item[(i)] $\partial \varphi \in [K_1, K_2]$,
\item[(ii)] $W_{\lambda}$ has no poles along $j\R$,
\item[(iii)] the Nyquist diagram of $W_{\lambda}$ encircles 
${(p-q)}$ times the point $-1/K_1$  in the clockwise direction, with $q$ the number of poles of $W_{\lambda}$ in $\Coplus$, and
\item[(iv)]  one of the following conditions holds
\begin{itemize}
\item[(a)] ${K_1 K_2 >0}$ and the Nyquist diagram of $W_{\lambda}$ lies \textit{outside} the disk $D(K_1, K_2)$,
\item[(b)] ${K_1 K_2 <0}$ and the Nyquist diagram of $W_{\lambda}$ lies \textit{inside} the disk $D(K_1, K_2)$,
\item[(c)] ${K_1 = 0}$ or ${K_2= 0}$ and the Nyquist diagram of $W_{\lambda}$ lies \textit{outside} the disk $D(K_1, K_2)$.
\end{itemize}
\end{itemize}
Then  system~\eqref{eq:system-Lure} is strictly $p$-dominant with rate $\lambda$.
\end{theorem}

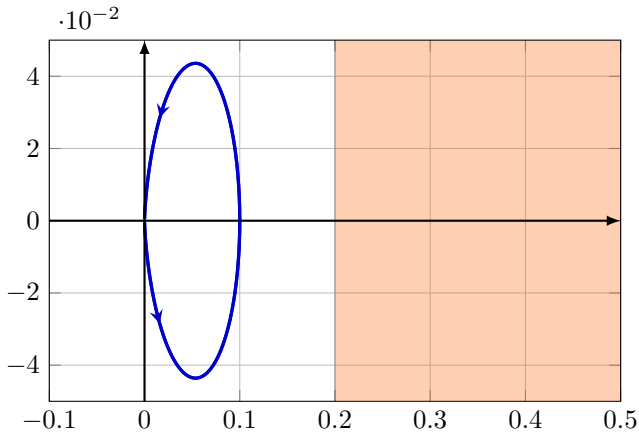
\begin{figure}[h!]
\centering
\pgfplotsset{compat=1.14}
\usetikzlibrary{decorations.markings}
\begin{tikzpicture}
\begin{axis}[
height = 0.35\textwidth,
width = 0.5\textwidth,
xmin=-0.1,xmax=0.5,
ymin=-0.05,ymax=0.05,
grid
]
\addplot [blue!80!black, very thick, %
postaction={decorate}, %
decoration={markings, mark=at position 0.15 with {\arrow{stealth}}} %
]  table [x=X, y=Y, col sep=comma]{msd21.csv};
\addplot [blue!80!black,  very  thick,  %
postaction={decorate}, %
decoration={markings, mark=at position 0.15 with {\arrowreversed{stealth}}} %
]   table [x=X, y=Y, col sep=comma]{msd22.csv};
\addplot[
		fill = orange!75!red, 
		opacity = 0.3,
		] 
		coordinates  {(0.2, -0.05)  (0.2, 0.05) (0.5, 0.05) (0.5, -0.05) };
\addplot[-latex, black,  thick,   domain=-0.05:0.05]  (0,{x}); 
\addplot[-latex, black, thick,   domain=-0.1:0.5 ]  ({x},0); 
\end{axis}
\end{tikzpicture}
\centering
\caption{The Nyquist diagram of the $\lambda$-shifted transfer function associated with~\eqref{eq:W-bar} (solid) lies outside the disk $D(-5,0)$ (shaded) for ${\lambda =2}$.  
} 
\label{fig:msd2}
\end{figure}

\begin{example}[The motivating example, continued]  \label{exa:case-study-1}
We use Theorem~\ref{thm:circle_criterion} to show that the closed-loop~\eqref{eq:system-MSD},~\eqref{eq:controller-P} is globally bistable for ${k_P = -2}$.  The system can be modelled as a Lure system of the form~\eqref{eq:system-Lure}, with  transfer function
\beq \label{eq:W}
W(s) =  \frac{1}{m s^2 + d s + k} ,
\eeq
and static nonlinearity  $\varphi: \R \to \R$ defined as
\beq  \label{eq:varphi}
\varphi (y) =   k_P \tanh(y) , \quad y \in \R,
\eeq 
which for ${k_P=-2}$ satisfies the differential sector condition ${\partial \varphi \in [K_1,K_2]}$, with ${K_1 = k_P}$ and ${K_2 = 0}$. Fig.~\ref{fig:msd1} shows that for  ${\lambda=2}$ the Nyquist diagram of the $\lambda$-shifted transfer function  $ W_\lambda $ (solid) lies outside the disk $D(-2,0)$ (shaded). Since $W_\lambda$ has one pole in $\Coplus$ and no poles along $j\R$,  Theorem~\ref{thm:circle_criterion} implies that system~\eqref{eq:system-MSD} is strictly $1$-dominant with rate ${\lambda =2}$.  By Theorem~\ref{thm:asymptotic}, every bounded solution of system~\eqref{eq:system-MSD}  must converge to an equilibrium point. In particular, \textit{all} solutions of the system must converge to an equilibrium point, since all solutions are bounded due to the bounded-input-bounded-output stability of both system~\eqref{eq:system-MSD} and the saturated nonlinearity~\eqref{eq:varphi}. Finally, a graphical argument shows that for ${k_P = -2}$ the system possesses three equilibria: the unstable equilibrium at the origin and two stable equilibria that are symmetric with respect to the origin. As a consequence, the system is globally bistable, as illustrated  by the numerical simulation  in Fig.~\ref{fig:msd_time_bistability}.

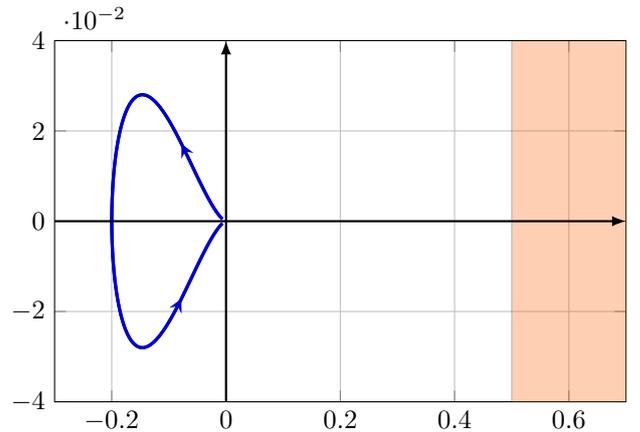
\begin{figure}[h!]
\centering
\pgfplotsset{compat=1.14}
\begin{tikzpicture}
\begin{axis}[
height = 0.35\textwidth,
width = 0.5\textwidth,
xmin=-0.3,xmax=0.7,
ymin=-0.04,ymax=0.04,
grid
]
\addplot [blue!80!black, very thick, %
postaction={decorate}, %
decoration={markings, mark=at position 0.15 with {\arrow{stealth}}} %
]  table [x=X, y=Y, col sep=comma]{msd11.csv};
\addplot [blue!80!black,  very  thick,  %
postaction={decorate}, %
decoration={markings, mark=at position 0.15 with {\arrowreversed{stealth}}} %
]   table [x=X, y=Y, col sep=comma]{msd12.csv};
\addplot[
		fill = orange!75!red, 
		opacity = 0.3,
		] 
		coordinates  {(0.5, -0.04)  (0.5, 0.04) (1.2, 0.04) (1.2, -0.04) }; 
\addplot[-latex, black,  thick,   domain=-0.04:0.04]  (0,{x}); 
\addplot[-latex, black, thick,   domain=-0.3:0.7 ]  ({x},0); 
\end{axis}
\end{tikzpicture}
\centering
\caption{The Nyquist diagram of the $\lambda$-shifted transfer function of system~\eqref{eq:system-MSD}  (solid) lies outside the disk $D(-2,0)$ (shaded) for ${\lambda =2}$.  
} 
\label{fig:msd1}
\end{figure}%

The discussion above illustrates a general principle. The combination of positive feedback and a saturated nonlinearity  is a basic mechanism to generate  multiple equilibria. Strict $1$-dominance guarantees convergence of all bounded solutions to one of those equilibria. Therefore this behavior   is robust if exogenous perturbations or parametric uncertainties preserve strict $1$-dominance and the presence of multiple equilibria. For example, for ${k_P < - 1}$ the closed-loop system possesses a single equilibrium at the origin. As a result, all solutions will converge to the unique equilibrium (monostability).

We conclude this example by observing that the same approach can be used to design a global oscillation using the feedback law~\eqref{eq:controller-PI}. This behavior can be generated through an additional integrator, as illustrated in Fig.~\ref{fig:integral}. 

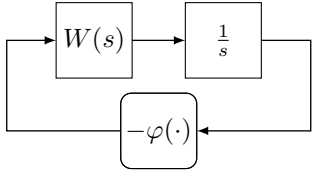
\begin{figure}[h!]
\centering 
\tikzstyle{sum} = [draw, circle, node distance=1cm]
\usetikzlibrary{arrows}
\begin{tikzpicture}
\draw  (-0.15,3.5) rectangle (0.85,2.5);
\node at (0.35,3) {$\frac{1}{s}$};
\draw[rounded corners]  (-1,2.3) rectangle (0,1.3);
\node at (-0.5,1.8) {$-\varphi(\cdot)$};
\draw[-latex, line width = .5 pt] (0.85,3) -- (1.5,3) -- (1.5,1.8) -- (0,1.8);
\draw[-latex, line width = .5 pt] (-1,1.8) -- (-2.5,1.8) -- (-2.5,3) --(-1.85,3) ;
\draw  (-1.85,3.5) rectangle (-0.85,2.5);
\node at (-1.35,3) {$W(s)$};
\draw[rounded corners]  (-1,2.3) rectangle (0,1.3);
\draw [-latex, line width = .5 pt](-0.85,3) -- (-0.15,3);
\end{tikzpicture}
\centering
\caption{The closed-loop system~\eqref{eq:system-MSD},~\eqref{eq:controller-PI}.}
\label{fig:integral}
\end{figure}%

Fig.~\ref{fig:msd2} shows that for ${\lambda = 2}$ the Nyquist diagram  of  the 
$\lambda$-shifted transfer function associated with
\beq  \label{eq:W-bar}
\overbar{W}(s)= \frac{ W(s)}{s},
\eeq 
lies to the left of the disk $D(-5,0)$. Since $\overbar{W}_{\lambda}$ has two poles in $\Coplus$ and no poles along $j\R$, Theorem~\ref{thm:circle_criterion} guarantees that the closed-loop system is strictly $2$-dominant with rate ${\lambda=2}$. Since for every ${k_I<0}$ the (unique) equilibrium point at the origin is unstable, the additional integral action produces indeed a global oscillation for ${k_I = -5}$, as illustrated in Fig.~\ref{fig:msd2_time_oscillations}. 
\hspace*{\fill} $\blacktriangle$
\end{example}

\section{Dominance margins} \label{sec:main-results}

This section introduces robustness margins for $p$-dominant systems, with the goal of characterizing quantitatively the robustness of multistable and oscillatory behaviors  in Lure systems. To this end, robustness of $p$-dominance is first studied in systems that can be described as the feedback interconnection of a linear, time-invariant system and a \textit{constant} gain. The analysis is subsequently extended to the wider class of Lure systems, where the feedback term can be a \textit{static, nonlinear} function.

\subsection{Gain and phase margins}

The Nyquist criterion criterion for $p$-dominance provides a necessary and sufficient condition for $p$-dominance of a linear, time-invariant system. In analogy with the Nyquist criterion for stability, this implicitly defines notions of gain and phase margins for $p$-dominance.

\begin{definition} \label{def:gain}
Let $W$ be the transfer function of a $p_1$-dominant linear, time-invariant system with rate ${\lambda \in \Rplus}$. An interval $(K_1, K_2) \subset \R$ is said to be a \textit{$p_2$-gain margin} with rate $\lambda$ if 
for every ${K \in (K_1,K_2)}$ the Nyquist diagram of the transfer function $W_{\lambda}$ encircles the point $-1/K$ $(p_2-p_1)$ times in the clockwise direction.
\end{definition}

\begin{definition} \label{def:phase}
Let $W$ be the transfer function of a $p_1$-dominant linear, time-invariant system with rate ${\lambda \in \Rplus}$. For ${K \in \R}$, with ${K \not = 0}$,   an interval $(\phi_1, \phi_2) \subset \R$ is said to be a  \textit{$p_2$-phase margin} with rate $\lambda$ if for every ${\phi \in (\phi_1, \phi_2)}$ a rotation of $\phi$ of the Nyquist diagram of the transfer function $W_{\lambda}$ encircles  the point $-1/K$ $(p_2-p_1)$ times in the clockwise direction.
\end{definition}

The $p$-gain margin measures how much the gain of the transfer function of a system can be increased before the closed-loop transfer function ceases to be  $p$-dominant. The $p$-phase margin measures how much phase lag can be added to the return ratio before the closed-loop system ceases to be $p$-dominant. Indeed, the notions of $0$-gain margin and $0$-phase margin with rate ${\lambda = 0}$ correspond to the classical notions of gain and phase margins given, \textit{e.g.}, in~\citep{sepulchre1997constructive}.

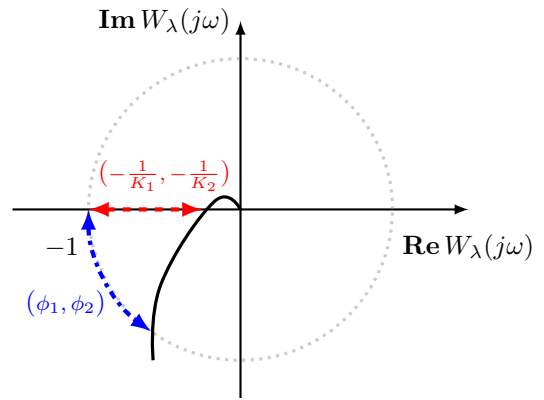
\begin{figure}[h!]
\centering 
\usetikzlibrary{decorations.markings}
\usetikzlibrary{patterns}
\begin{tikzpicture}[scale=1, every node/.style={transform shape}]
\draw [thick, -latex](9.5,-0.5) -- (9.5,4.5);
\draw [thick, -latex](6.5,2) -- (12.5,2);
\draw[very thick, dotted, opacity = 0.4, color = gray]  (9.5,2) ellipse (2 and 2);
\draw[very thick]  plot[smooth , tension=.7] coordinates {(9.5,2)  (9.15,2.1) (8.45,1)(8.35,0)};
\node at (12.5,1.5) {$\mathbf{Re} \,W_{\lambda}(j\omega)$};
\node at (8.5,4.5) {$\mathbf{Im} \,W_{\lambda}(j\omega)$};
\node[red] at (8.5,2.45) {({\footnotesize $-\frac{1}{K_1},-\frac{1}{K_2}$})};
\node[blue] at (7.2,0.75) {({\footnotesize $\phi_{1},\phi_{2}$})};
\node at (7.15,1.5) {$-1$};
\draw[latex-latex, ultra thick, dashdotted, color = blue] (7.5,2) arc (180:234:2);
\draw[latex-latex, ultra thick, dashed, color = red]  (7.5,2) -- (9,2);
\end{tikzpicture}
\centering
\caption{Diagrammatic illustration of the notions of $p$-gain margin (dashed) and $p$-phase margin (dashdotted).}
\label{fig:gain_phase_margin}
\end{figure}

In analogy with classical gain and phase margins, $p$-gain and $p$-phase margins of a system can be obtained directly from the number of encirclements of Nyquist diagram of the  $\lambda$-shifted transfer function $W_\lambda$ around the critical point $-1/K$. If the Nyquist diagram of $W_\lambda$ does not encircle  the critical point $-1/K$, the $p$-dominance properties of the closed-loop system  are unchanged (${p_1=p_2}$). By contrast, if the Nyquist diagram of $W_\lambda$ encircles the critical point $-1/K$ $n_e$ times in the clockwise direction, then the closed-loop system is strictly ${(p_1+n_e)}$-dominant with rate $\lambda$. Fig.~\ref{fig:gain_phase_margin} provides a diagrammatic illustration of the notions of $p$-gain margin and $p$-phase margin.

\begin{figure}[h!]
\centering 
\definecolor{purple1}{rgb}{0.69, 0, 0.98}
\pgfplotsset{%
  colormap={whitered}{color(0cm)=(white);  color(1cm)=(orange!75!red)}
}
\begin{tikzpicture}[scale=0.525, every node/.style={transform shape},
    declare function = {margin = -x^2 + y*x-1; },
    declare function = {K_opt = 0.25*y^2 -1; },
  ]
  \begin{axis}[
    colormap name  = whitered,
    width          = 15cm,
    view           = {40}{60},
    enlargelimits  = false,
    grid           = major,
    domain         = 0:10,
    y domain       = 0:10,
    xlabel         =  {\huge ${\lambda}$}, 
    ylabel         = {\huge $d$},  
    zlabel         = {\huge  $K$ },  
    z label style= {
    		at  = {(0.1,.85)}, 
    		rotate=-90,
    		anchor=north
    	},
    ticklabel style = {font=\Large }
  ]
\addplot3 [ surf, opacity = 0.7 ]{margin};   
\addplot3[ smooth, line width= 3.25pt, color = purple1]  (0.5*y,y,{K_opt} );  
  \end{axis}
\end{tikzpicture}
\centering
\caption{The $1$-gain margin of system~\eqref{eq:system-MSD} as a function of the rate $\lambda$ and the damping coefficient $d$, with ${m = 1 \, \mathrm{kg}}$ and ${k = 1 \,\mathrm{N} \cdot \mathrm{m}^{-1}}$.
The solid line describes the optimal rate $\lambda$ for a given damping coefficient $d$.}
\label{fig:margins3d}
\end{figure}
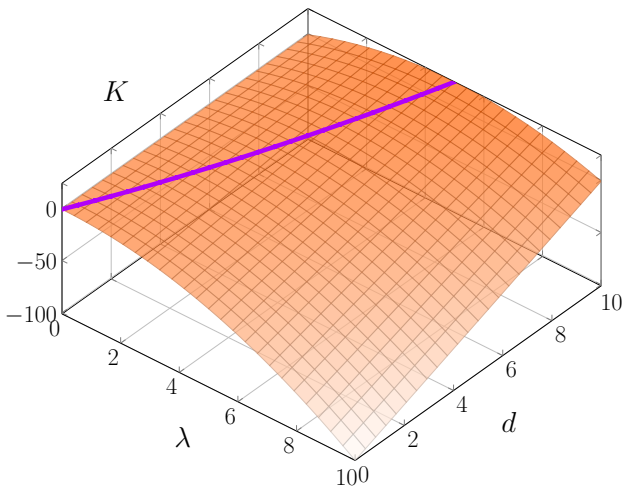

\begin{example}[The motivating example, continued]  \label{exa:case-study-2} 
Consi-\linebreak\-der the mass-spring-damper system~\eqref{eq:system-MSD}. Fig.~\ref{fig:msd1} shows that 
$(-\infty,5)$ is a $1$-gain margin with rate ${\lambda = 2}$, while $(5,\infty)$ is a $0$-gain margin with the same rate, respectively. For ${\lambda = 2}$ the Nyquist diagram of the transfer function $W_{\lambda}$ (solid)  does not encircle $-1/K$ for every  ${K \in (-\infty,5)}$ and encircles $-1/K$ once in the anticlockwise direction for every ${K \in (5,\infty)}$. As a result, the closed-loop system is strictly $0$-dominant with rate ${\lambda = 2}$ for  $K\in (5,\infty)$ and  strictly $1$-dominant with the same rate for $K \in (-\infty,5)$, respectively. Observe that an increase in the damping coefficient results in a larger separation of the poles of the transfer function~\eqref{eq:W}, which  raises the question of the influence of the rate $\lambda$ on dominance margins. Fig.~\ref{fig:margins3d} shows the  $1$-gain margin of system~\eqref{eq:system-MSD} as a function of the rate $\lambda$ and the damping coefficient $d$, with ${m = 1 \, \mathrm{kg}}$ and ${k = 1 \,\mathrm{N} \cdot \mathrm{m}^{-1}}$. The solid line  describes the optimal rate $\lambda$ for a given damping coefficient $d$. Interestingly, to obtain the same $1$-gain margin one needs to increase the rate $\lambda$ as the damping coefficient $d$ grows. 
\hspace*{\fill} $\blacktriangle$
\end{example}

\subsection{Disk margins}

We have seen that $p$-gain and $p$-phase margins are important indicators of the robustness of $p$-dominance in feedback \textit{linear} systems. We now introduce the notion of $p$-disk margin to measure the robustness of $p$-dominance in Lure systems, which can give rise to a broader spectrum of behaviors due to the possibly \textit{nonlinear} feedback term. 

\begin{definition} \label{def:disk}
Let $W$ be the transfer function of a $p_1$-dominant linear, time-invariant system with rate ${\lambda \in \Rplus}$. The transfer function $W$ is said to have a \textit{$p_2$-disk margin $D(K_1,K_2)$}, with at most one of $K_1$ and $K_2$ nonzero, with rate $\lambda$ if the Nyquist diagram of the transfer function $W_{\lambda}$ does not intersect the disk  $D(K_1,K_2)$ and  encircles it  $(p_2-p_1)$ times in the clockwise direction.
\end{definition}

For linear systems robustness margins are defined in terms of the number of encirclements around a critical point. Intuitively, for Lure systems the critical point becomes a disk and robustness is measured in terms of the number of encirclements around the ``critical disk''. This intuition is formalised by the following statement, where the circle criterion for $p$-dominance is reformulated using the notion of $p$-disk margin.

\begin{theorem}\label{thm:circle_criterion_disk}
Let $W$ be the transfer function of a $p_1$-dominant linear, time-invariant system with rate ${\lambda \in \Rplus}$. If $W$ has a $p_2$-disk margin $D(K_1,K_2)$, with at most one of $K_1$ and $K_2$ nonzero, with rate $\lambda$ then the negative feedback interconnection of $W$ and the function $\varphi$ is $p_2$-dominant for every $\varphi$ such that $\partial \varphi \in [K_1, K_2]$.
\end{theorem}

The notion of $p$-disk margin is stronger than of those of $p$-gain and  $p$-phase   margins. Assuming for simplicity that ${K_1K_2>0}$, if $W$ has $p_2$-disk margin $D(K_1,K_2)$ with rate ${\lambda}$, then it has a $p_2$-gain margin $(K_1,K_2)$ with the same rate, since for every ${K \in (K_1,K_2)}$ the point $-1/K$ is in the interior of the disk $D(K_1,K_2)$ and the Nyquist diagram of the transfer function $W_{\lambda}$ satisfies the encirclement condition. Similarly, if ${K^{\star} 
\in \R}$ is such that ${-1/K^{\star} \in D(K_1,K_2)}$ then a graphical argument shows that the $p_2$-phase margin is  at least a given interval $(\phi_{1}, \phi_{2})$.

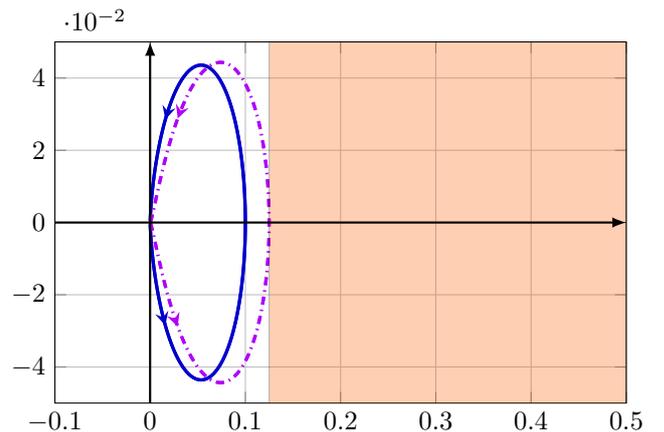
\begin{figure}[h!]
\centering
\definecolor{purple1}{rgb}{0.69, 0, 0.98}
\definecolor{blue1}{rgb}{0, 0.72, 0.98}

\pgfplotsset{compat=1.14}
\usetikzlibrary{decorations.markings}
\begin{tikzpicture}
\begin{axis}[
height = 0.35\textwidth,
width = 0.5\textwidth,
xmin=-0.1,xmax=0.5,
ymin=-0.05,ymax=0.05,
grid
]
\addplot [blue!80!black, very thick, %
postaction={decorate}, %
decoration={markings, mark=at position 0.15 with {\arrow{stealth}}} %
]  table [x=X, y=Y, col sep=comma]{msd21.csv};
\addplot [blue!80!black,  very  thick,  %
postaction={decorate}, %
decoration={markings, mark=at position 0.15 with {\arrowreversed{stealth}}} %
]   table [x=X, y=Y, col sep=comma]{msd22.csv};
\addplot [color = purple1, dashdotted, very thick, %
postaction={decorate}, %
decoration={markings, mark=at position 0.15 with {\arrow{stealth}}} %
]  table [x=X, y=Y, col sep=comma]{msd3101.csv};
\addplot [color = purple1, dashdotted,  very  thick,  %
postaction={decorate}, %
decoration={markings, mark=at position 0.15 with {\arrowreversed{stealth}}} %
]   table [x=X, y=Y, col sep=comma]{msd3201.csv};
\addplot[
		fill = orange!75!red, 
		opacity = 0.3,
		] 
		coordinates  {(0.125, -0.05)  (0.125, 0.05) (0.5, 0.05) (0.5, -0.05) };
\addplot[-latex, black,  thick,   domain=-0.05:0.05]  (0,{x}); 
\addplot[-latex, black, thick,   domain=-0.1:0.5 ]  ({x},0); 
\end{axis}
\end{tikzpicture}
\centering
\caption{The Nyquist diagram of the $\lambda$-shifted transfer function associated with~\eqref{eq:W-bar} with (dashdotted) and without (solid) actuator dynamics lies outside the disk $D(k_I,0)$ (shaded) for  ${k_I \in (-8,0)}$ and ${\lambda =2}$.  
} 
\label{fig:msd3}
\end{figure}

\begin{example}[The motivating example, continued]  \label{exa:case-study-3} 
We now model the effect of actuator dynamics in the design of a robust global oscillation. The transfer function
\beq \label{eq:W_actuator}
C(s) = \frac{1}{\tau s + 1}, \quad \tau \in \Rplus,
\eeq
is added in series to $\overbar{W}$ in the block diagram in Fig.~\ref{fig:integral}.
Fig.~\ref{fig:msd3} shows that the Nyquist diagram of the $\lambda$-shifted transfer function associated with~\eqref{eq:W-bar} with (dashdotted) and without (solid) actuator dynamics lies outside the disk $D(k_I,0)$ (shaded) for every ${k_I \in (-8,0)}$ for ${\lambda =2}$. As a result, the disk $D(k_I,0)$ is a $2$-disk margin with rate ${\lambda = 2}$ for every ${k_I \in (-8,0)}$ for system~\eqref{eq:system-MSD},~\eqref{eq:controller-PI} when the actuator dynamics is taken into account. By Theorem~\ref{thm:circle_criterion_disk}, the closed-loop system is strictly $2$-dominant with rate ${\lambda=2}$ and produces a global oscillation for every ${k_I \in (-8,0)}$. This means, for example, that a variation that is strictly less than $\pm  35 \%$ from the nominal gain ${k_I=-5}$ of the controller~\eqref{eq:controller-PI} still produces a globally oscillating behavior. 
\hspace*{\fill} $\blacktriangle$
\end{example}

\section{Discussion}  \label{sec:discussion}

The present paper can be considered as a first step towards a quantitative and tractable theory of robustness of systems that switch and oscillate. The proposed framework has been developed following the frequency domain approach to $p$-dominance theory in~\citep{felix2018analysis}. The presence of a Nyquist criterion for $p$-dominance has led to the definition of dominance margins, which provide intuitive measures of how the dominant poles of system affect the closed-loop behavior. The theory developed can extended in several ways.  

A promising research direction is that of revisiting $H_{\infty}$ control theory in the light of the notion of $p$-dominance. This can be intuitively seen from the fact that stability margins are intimately connected to the $H_{\infty}$ norm. In the same way, connections can be established between $p$-dominance margins with respect to the rate ${\lambda \in \Rplus}$ and the norm
\beq \label{eq:H_inf_norm}
\norm{W}_{{\infty,\lambda}} = \mathrm{ess} \sup_{\omega \in \R} |W(j\omega - \lambda)| .
\eeq

The definition of an $H_{\infty}$ norm, in turn,  naturally leads to the development of new frequency domain tools geared towards the analysis of multistable and oscillatory Lure systems. For analysis purposes, one may establish robust versions of the Nyquist criterion and the circle criterion for $p$-dominance using small gain conditions with respect to the norm~\eqref{eq:H_inf_norm}. Different uncertainty models could be used, including additive perturbations, multiplicative perturbations and coprime factor perturbations~\citep{doyle1992feedback,zhou1996robust}.  For design purposes, a \textit{loop shaping} technique could be devised using the norm~\eqref{eq:H_inf_norm}. These tools are envisioned as bridging factors between theory and practice in the analysis and design of switches and oscillators.

\begin{figure}[h]
\centering 
\usetikzlibrary{decorations.markings}
\usetikzlibrary{patterns}
\definecolor{blue1}{rgb}{0, 0.72, 0.98}
\begin{tikzpicture}[scale=1, every node/.style={transform shape}]
\draw [-latex](-0.5,22.5) -- (-0.5,26);
\draw [-latex](-0.5,24.5) -- (6.5,24.5);
\draw (-0.5,23) -- (2.5,26);
\draw (1,26) -- (1,22.5);
\node at (7,24.5) {$\lambda$};
\node at (3,25) {$\lambda =\alpha$};
\node at (-1,26) {$K$};
\node at (-1,23) {$\frac{1}{\alpha}$}; 
\fill[color=orange!75!red, opacity = 0.3] (-0.5,24.5) -- (-0.5,26) -- (1,26) -- (1,24.5) -- cycle;
\fill[color=orange!75!red, opacity = 0.3] (6.5,26) -- (2.5,26) -- (1,24.5) --  (6.5,24.5) -- cycle;
\fill[color=orange!75!red, opacity = 0.3] (1,22.5) -- (-0.5,22.5) -- (-0.5,23)-- (1,24.5) -- cycle;
\fill[color=blue1, opacity = 0.15] (1,22.5) -- (1,24.5) -- (6.5,24.5) -- (6.5,22.5) -- cycle;
\fill[color=blue1, opacity = 0.15] (1,24.5)  -- (1,26) -- (2.5,26) -- cycle;
\fill[color=blue1, opacity = 0.15] (-0.5,23) -- (-0.5,24.5) -- (1,24.5) -- cycle;
\fill[pattern=dots] (1,22.5) -- (1,24.5) -- (6.5,24.5) -- (6.5,22.5) -- cycle;
\fill[pattern=dots] (1,24.5)  -- (1,26) -- (2.5,26) -- cycle;
\fill[pattern=dots] (-0.5,23) -- (-0.5,24.5) -- (1,24.5) -- cycle;
\draw [-latex] (2.45,24.95) -- (1.35,24.6);
\end{tikzpicture}
\centering
\caption{The $0$-gain margins (dotted) and $1$-gain margins (shaded) of~\eqref{eq:W_example1} as functions of the rate $\lambda$.}
\label{fig:margins_lambda}
\end{figure}
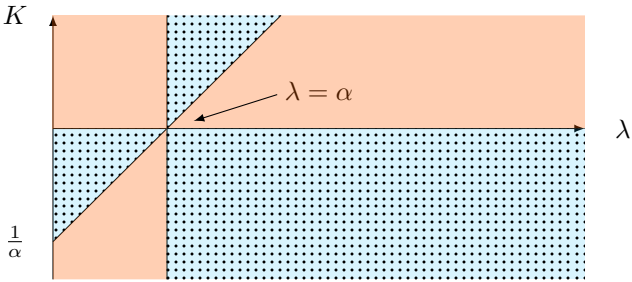

It should be noted that the $p$-dominance properties of a system depend on the rate $\lambda$. For instance, a system may possess different $p$-dominance margins for different values of $\lambda$. Fig.~\ref{fig:margins_lambda} illustrates   an   example of this fact showing the $0$-gain and the $1$-gain margins of the transfer function
\beq \label{eq:W_example1}
W(s) = \frac{1}{(s+\alpha)}, \quad \alpha \in \Rplus,
\eeq
as functions of the rate $\lambda$. This means that the rate $\lambda$ can be regarded as a parameter to be optimized. Clearly, the further from a singularity, the larger the robustness margins. 

Another important research direction consists in revisiting our results in the light of \textit{$p$-dissipativity theory}~\citep{forni2018differential}.  An interconnection theory can be built using \textit{small $p$-gain conditions} and  a  robust control framework can be developed  for general nonlinear systems that exhibit multistable and oscillatory behaviors.  At a computational level, \textit{linear matrix inequalities} with inertia constraints are expected to play an important role. 

Finally, more complex case studies need to be considered to assess the engineering relevance of the theory developed.

\section{Conclusion} \label{sec:conclusion}

The paper has introduced notions of robustness margins tailored to the analysis and design of systems that switch and oscillate. Dominance margins have been defined as a natural generalization of stability margins to measure the robustness of $p$-dominance in a system. These notions have been shown to possess nice graphical interpretations in the frequency domain and to be fruitful tools for the analysis and design of multistability  and oscillations in Lure systems. The theory has been illustrated by means of a simple mechanical example.

\end{document}